\documentclass[%
reprint,
superscriptaddress,
 amsmath,amssymb,
 aps,
prb,
floatfix,
]{revtex4-1}

\usepackage{graphicx}
\usepackage{dcolumn}
\usepackage{bm}
\usepackage[bookmarks=true,colorlinks=true,linkcolor={blue},citecolor=blue,urlcolor=blue]{hyperref} 
\usepackage[mathlines]{lineno}

\usepackage{longtable}	
\usepackage{tabularx}	
\usepackage{array}		
\usepackage{multirow}	
\usepackage{array}		
\usepackage{longtable}
\begin{document}


\title{Computational study of the structural, electronic and optical properties of bulk palladium nitrides}


\author{Mohammed S. H. Suleiman}
\email[Corresponding author: ]{suleiman@aims.ac.za}
\affiliation{School of Physics, University of the Witwatersrand, Johannesburg, South Africa.}
\affiliation{Department of Physics, Sudan University of Science and Technology, Khartoum, Sudan.}

\author{Daniel P. Joubert}
\homepage[Homepage: ]{http://www.wits.ac.za/staff/daniel.joubert2.htm}
\affiliation{School of Physics, University of the Witwatersrand, Johannesburg, South Africa.}

\date{\today}
\begin{abstract}
The atomic and electronic structures of Pd$_3$N, PdN and PdN$_{2}$ were investigated using \textit{ab initio} density-functional theory (DFT). We studied cohesive energy \textit{vs.} volume equation of states (EOS) for a set of reported and hypothetical structures. Obtained data was fitted to a third-order Birch-Murnaghan equation of state (EOS) so as to identify the energetically most stable phases and to determine their equilibrium structural parameters and stability and mechanical properties. Electronic properties were investigated by calculating the band diagrams and the total and partial density of states (DOS). Some possible pressure-induced phase transitions were tested. To derive the frequency-dependent optical spectra (i.e. absorption coefficient, reflectivity, refractive index, and energy-loss), we performed $G_0W_{0}$ calculations within the random-phase approximation (RPA) to the dielectric tensor. Obtained results were compared with previous studies.
\end{abstract}
%
\pacs{}

\maketitle

\tableofcontents	
\section{\label{Introduction}Introduction}
In 2007, Crowhurst \textit{et al.} \cite{PdN2_2008_exp_n_comp} reported the synthesis of the new palladium nitride compound and argued for its PdN$_{2}$ stoichiometry and pyrite (C2) structure. However, many transition-metal nitrides (TMNs) are known to form more than one nitride \cite{StructuralInChem}, and first-principles methods are commonly employed to search for possible stable phases.

In the present work, we consider Pd$_3$N, PdN and PdN$_{2}$ stoichiometries in $20$ different crystal structures. Equation of states (EOS), structural preferences and thermodynamic stabilities for these three stoichiometric series are analyzed and the equilibrium lattice parameters are determined. Some possible pressure-induced phase transitions are tested. The band-structure and the density of states (DOS) of the relatively most stable modifications are carefully investigated. Obtained results are compared with previous calculations and with experiment. We furthermore present the electronic structure and the GWA-derived frequency-dependent optical constants of a high-pressure favored PdN candidate, PdN(B24).

This article is organized as follows: the methods of calculations is given in Section \ref{Calculation Methods}. Obtained results are presented and discussed in Section \ref{Results and Discussion}. The article is concluded with some remarks in Section \ref{Concluding Remarks}.
%
\section{\label{Calculation Methods}Calculation Methods}
%
\subsection{\label{Stoichiometries and Crystal Structures} Stoichiometries and Crystal Structures}
Considered crystal structures are given in Table \ref{PdN's allstructures}. In this table, Pd$_m$N$_n$ modifications are divided, according to the nitrogen content, into four series with $ \frac{n}{m}  = 0, \frac{1}{3}, 1 \text{ and } 2$. Within each series, structures are then arranged in a descending structural symmetry order, that is from the highest space group number to the least one.

\begin{table}[h]
\caption{The studied structural modifications of Pd, Pd${_3}$N, PdN and PdN${_2}$. Presented information are the Strukturbericht designation (symbol), prototype compounds, the sequential number of the space group (\#SG) as given in the \href{http://it.iucr.org/A/}{International Tables for Crystallography}, and the number of Pd${_\text{m}}$N${_\text{n}}$ formulas per unit cell ($Z$).}
\begin{tabular}{l|lll}
\hline \hline	
Symbol			&Prototype(s)						&\#SG			&$Z$      \\
\hline
\multicolumn{4}{c}{\textbf{Pd Structure}} \\
\hline	
A1              &Cu									&225   		& 1       \\

\hline	
\multicolumn{4}{c}{\textbf{Pd$_3$N Structures}} \\
\hline	
D0$_3$          &AlFe$_3$							&225 		&1 		\\

A15             &Cr$_3$Si 							&223		&2     	\\

D0$_9$          &anti-ReO$_3$ ($\alpha$), Cu$_3$N	&221		&1     \\

L1$_2$          &Cu$_3$Au                   		&221		&1     	\\

D0$_2$          &CoAs$_3$ (skutterudite)    		&204   		&4     	\\

$\epsilon$-Fe$_3$N&$\epsilon$-Fe$_3$N, Ni$_3$N 		&182        &2     \\

RhF$_3$         &RhF$_3$                    		&167   		&2     	\\

\hline	
\multicolumn{4}{c}{\textbf{PdN Structures}} \\
\hline	
B1              &NaCl								&225   		& 1       \\

B2              &CsCl								&221   		& 1       \\  

B3              &ZnS (zincblende)					&216   		& 1       \\   

B8$_{1}$        &NiAs								&194   		& 2       \\

B$_{\text{k}}$  &BN									&194	   	& 2       \\

B$_{\text{h}}$  &WC									&187 		& 1       \\

B4              &ZnS (wurtzite)						&186	    & 2       \\

B17             &PtS (cooperite)					&131 	    & 2       \\

B24             &TlF								&69         & 1       \\

\hline	
\multicolumn{4}{c}{\textbf{PdN$_2$ Structures}} \\
\hline	
C1				&CaF$_{2}$ (fluorite)				&225   		& 1       \\

C2				&FeS$_{2}$ (pyrite)					&205	    & 4       \\  

C18				&FeS$_{2}$ (marcasite)				&58			& 2       \\   

CoSb$_{2}$		&CoSb$_{2}$	 						&14			& 4       \\

\hline\hline
\end{tabular}
\label{PdN's allstructures}
\end{table}

It is interesting to investigate the possibility of formation of Pd$_3$N. To the best of our knowledge, this stoichiometry has not been considered before for the Pd-N system, though Ni, which is the first element of group 10 in the periodic table, is known to form Ni$_3$N in the hexagonal $\epsilon$-Fe$_3$N structure \cite{Ni3N_structure}.
%
%
\subsection{\label{Electronic Relaxation Details}Electronic Relaxation Details}
In the present investigation, electronic structure spin density functional theory (SDFT) \cite{SDFT_1972,SDFT_Pant_1972} calculations have been employed using the VASP code\cite{Vasp_ref_PhysRevB.47.558_1993,Vasp_ref_PhysRevB.49.14251_1994,Vasp_cite_Kressw_1996,
Vasp_PWs_Kresse_1996,DFT_VASP_Hafner_2008,PAW_Kresse_n_Joubert}.

VASP implements the projector augmented wave (PAW) method\cite{PAW_Blochl, PAW_Kresse_n_Joubert} to describe the core-valence interactions $V_{\text{ext}}(\mathbf{r})$, where the $4d^{9}5s^{1}$ electrons of Pd and the $2s^{2}2p^{3}$ electrons of N are treated explicitly as valence electrons. The PAW potential treats the core electrons in a fully relativistic fashion, while for these valence electrons only scalar kinematic relativistic effects are incorporated \cite{DFT_VASP_Hafner_2008}.

VASP self-consistently solves the Kohn-Sham (KS) equations \cite{KS_1965}
\begin{eqnarray}	\label{KS equations}
\begin{split}
  \Bigg \{ - \frac{\hbar^{2}} {2m_{e}}  \nabla^{2} + \int d\mathbf{r}^{\prime} \frac{n(\mathbf{r}^{\prime})}{|\mathbf{r}-\mathbf{r}^{\prime}|} + V_{\text{ext}}(\mathbf{r}) \\  + V_{XC}^{\sigma, \mathbf{k}}[n(\mathbf{r})] \Bigg \} \psi_{i}^{\sigma, \mathbf{k}}(\mathbf{r})  =  
   \epsilon_{i}^{\sigma, \mathbf{k}} \psi_{i}^{\sigma, \mathbf{k}}(\mathbf{r}),
\end{split}
\end{eqnarray}
where $i$, $\mathbf{k}$ and $\sigma$ are the band, $\mathbf{k}$-point and spin indices, receptively, by expanding the pseudo part of the KS one-electron spin orbitals $\psi_{i}^{\sigma , \mathbf{k}}(\mathbf{r})$ on a basis set of plane-waves (PWs). Using PWs cut-off energy $E_{\text{cut}} = 600 \; eV$ and $\mathbf{\Gamma}$-centered Monkhorst-Pack \cite{MP_k_mesh_1976} $17 \times 17 \times 17$ meshes, the total energy converges to $< 3 \; \text{m} eV/ \text{atom}$.

While partial occupancies were set using the smearing method of Methfessel-Paxton (MP) \cite{MP_smearing_1989} in the ionic relaxation, the tetrahedron method with Bl\"{o}chl corrections \cite{tetrahedron_method_theory_1971,tetrahedron_method_theory_1972,ISMEAR5_1994} was used in the static calculations.

The implemented generalized gradient approximation (GGA), as parametrized by Perdew, Burke and Ernzerhof (PBE) \cite{PBE_GGA_1996,PBE_GGA_Erratum_1997,XC_PBE_1999} was employed for the exchange-correlation potentials $V_{XC}^{\sigma, \mathbf{k}}[n(\mathbf{r})]$.
%
\subsection{\label{Geometry Relaxation and EOS}Geometry Relaxation and EOS} 
Following the VASP implemented conjugate-gradient (CG) algorithm, ions which possess free internal parameters were relaxed until all force components on each ion were $< 0.01 \; eV/\text{\AA}$. This was done at a set of externally imposed lattice constants. To obtain very accurate and unambiguous total energies, one additional static calculation (as described in Subsection \ref{Electronic Relaxation Details}) at the end of each ion relaxation step.

The cohesive energy per atom $E_{\text{coh}}$ was calculated from
\begin{eqnarray} \label{E_coh equation}
E_{\text{coh}}^{\text{Pd}_{m}\text{N}_{n}}  =   \frac{  E_{\text{solid}}^{\text{Pd}_{m}\text{N}_{n}} - Z \times \left( m E_{\text{atom}}^{\text{Pd}} + n E_{\text{atom}}^{\text{N}} \right) }{Z \times (m + n)} \; ;
\end{eqnarray}
where $Z$ is the number of Pd$_{m}$N$_{n}$ formulas per unit cell, $E_{\text{atom}}^{\text{Pd}}$ and $E_{\text{atom}}^{\text{N}}$ are the energies of the non-spherical spin-polarized isolated pseudo-atom, and $m,n = 1,2 \text{ or } 3$ are the stoichiometric weights. We then fitted \footnote{Using J. K. Dewhurst's
Equation of State (EOS) program for fitting energy-volume data, EOS version 1.2, August 2005, \url{http://elk.sourceforge.net}.} the obtained $E_{\text{coh}}$ of the relaxed ions as a function of volume per atom $V$ to a Birch-Murnaghan 3rd-order equation of state (EOS)\cite{BM_3rd_eos}. From the fitting, the equilibrium volume $V_{0}$, the equilibrium cohesive energy $E_{0}$, the equilibrium bulk modulus
\begin{equation}	\label{B_0 eq}
B_{0} = -V \frac{\partial P}{\partial V}\Big|_{V=V_{0}} = -V \frac{\partial^{2} E}{\partial V^{2}}\Big|_{V=V_{0}}
\end{equation}
and its pressure derivative
\begin{equation}	\label{B^prime eq}
 B^{\prime}_{0} = \frac{\partial B}{\partial P} \Big|_{P=0} = \frac{1}{B_{0}} \left(  V \frac{\partial}{\partial V} (V \frac{\partial^{2} E}{\partial V^{2}}) \right)  \Big|_{V=V_{0}} 	
\end{equation}
were determined.
%
\subsection{\label{Formation Energy Calculations}Formation Energy Calculations}
Beside the cohesive energy, another measure of relative stability is the formation energy $E_{\text{f}}$. Assuming that palladium nitrides Pd$_m$N$_n$ result from the interaction between the N$_{2}$ gas and the solid Pd(A1) via the reaction
\begin{eqnarray} \label{E_f reaction}
m \text{Pd}^{\text{solid}} + \frac{n}{2} \text{N}_2^{\text{gas}} \rightleftharpoons   \text{Pd}_m\text{N}_n^{\text{solid}},
\end{eqnarray}
$E_{\text{f}}$ can be given by
\begin{align} \label{formation energy equation}
E_{\text{f}}(\text{Pd}_m\text{N}_n^{\text{solid}}) =   E_\text{coh}(\text{Pd}_m\text{N}_n^{\text{solid}}) \quad \quad \quad \quad \quad \quad &
\nonumber \\
- \frac{  m E_\text{coh}(\text{Pd}^{\text{solid}}) + \frac{n}{2} E_\text{coh}(\text{N}_2^{\text{gas}})}{m + n} &		\; .
\end{align}

Here $m,n=1,2,3$ are the stoichiometric weights and $E_\text{coh}(\text{Pd}_m\text{N}_n^{\text{solid}})$ is the cohesive energy per atom as in Eq. \ref{E_coh equation}. The obtained cohesive energy $E_\text{coh}(\text{Pd}^{\text{solid}})$ and the other equilibrium properties of the elemental metallic palladium are given in Table \ref{table: palladium_nitrides_equilibrium_structural_properties}. We found (Ref. \onlinecite{Suleiman_PhD_arxiv_copper_nitrides_article}) the cohesive energy of the diatomic nitrogen $E_\text{coh}(\text{N}_2^{\text{gas}})$ to be $-5.196 \; eV/\text{atom}$ for an equilibrium N--N bond length of $1.113 \; \text{\AA}$.
%
\subsection{\label{GWA Calculations}GWA Calculations}
In practice, one may get a qualitative agreement between DFT-calculated optical properties and experiment. However, technically speaking, neither KS eigenvalues correspond to true electron removal or addition energies nor their differences correspond to optical (neutral) excitations. To achieve accurate quantitative description of electronic excitations, one needs to go beyond the level of DFT \cite{PAW_optics,Richard_Martin,DFT_vs_GW_2002}.\\

Many-body perturbation theory (MBPT) provides an approach that leads to a system of equations, called  quasi-particle (QP) equations, which can be written for a periodic crystal as \cite{GWA_and_QP_review_1999,Kohanoff,JudithThesis2008}
\begin{eqnarray}	\label{QP equations}
\begin{split}
  \Bigg \{ - \frac{\hbar^{2}} {2m}  \nabla^{2} + \int d\mathbf{r}^{\prime} \frac{n(\mathbf{r}^{\prime})}{|\mathbf{r}-\mathbf{r}^{\prime}|} + V_{\text{ext}}(\mathbf{r}) \Bigg \} \psi_{i,\mathbf{k}}^{\text{QP}}(\mathbf{r}) \\  + \int d\mathbf{r}^{\prime} \Sigma(\mathbf{r},\mathbf{r}^{\prime};\epsilon_{i,\mathbf{k}}^{\text{QP}})  \psi_{i,\mathbf{k}}^{\text{QP}}(\mathbf{r}^{\prime}) = \epsilon_{i,\mathbf{k}}^{\text{QP}} \psi_{i,\mathbf{k}}^{\text{QP}}(\mathbf{r}).
\end{split}
\end{eqnarray}
The QP orbitals $\psi_{i,\mathbf{k}}^{\text{QP}}(\mathbf{r})$ were taken from the DFT calculations. However, in consideration of computational cost, we used a less dense $\mathbf{k}$-mesh of $4 \times 4 \times 4$. The so-called self-energy $\Sigma(\mathbf{r},\mathbf{r}^{\prime};\epsilon_{i,\mathbf{k}}^{\text{QP}})$ contains all the static and dynamic exchange-correlation effects, including those neglected in our KS-GGA unperturbed system. In the well-known $GW$ approximation \cite{Hedin_1st_GWA_1965}, $\Sigma$ is given in terms of the the one-particle Green's function $G$ of the many-body system as
\begin{eqnarray}	\label{GW self-energy}
\begin{split}
\Sigma_{GW} = j \int d\epsilon^{\prime} G(\mathbf{r},\mathbf{r}^{\prime};\epsilon,\epsilon^{\prime}) W(\mathbf{r},\mathbf{r}^{\prime};\epsilon)	\; .
\end{split}
\end{eqnarray}
The bare Coulomb interaction $v$ and the dynamically (i.e. frequency-dependent) screened Coulomb interaction $W$ are related via
\begin{eqnarray}
\begin{split}
W(\mathbf{r},\mathbf{r}^{\prime};\epsilon) = j \int d\mathbf{r}_{1} \varepsilon^{-1}(\mathbf{r},\mathbf{r}_{1};\epsilon)v(\mathbf{r}_{1},\mathbf{r}^{\prime}),
\end{split}
\end{eqnarray}
with $\varepsilon$, the dielectric matrix, calculated within the so-called random phase approximation (RPA). 

In the present study, we employed the $G_0W_0$ (i.e. single shot $GW$) routine in which the QP eigenvalues\cite{Kohanoff,JudithThesis2008,VASPguide}
\begin{eqnarray} \label{QP eigenvalues}
\begin{split}
\epsilon_{i,\mathbf{k}}^{\text{QP}} = \text{Re}  \left( 
\left\langle \psi_{i,\mathbf{k}}^{\text{QP}} \middle| 
H_{\text{KS}} - V_{XC} + \Sigma_{GW_{0}}
\middle|  \psi_{i,\mathbf{k}}^{\text{QP}} \right\rangle  	\right)
\end{split}
\end{eqnarray}
are updated only once in the calculations of $G$, while $W$ is kept at the DFT-RPA level. As implemented in VASP, the head and the wings of $\varepsilon$ are constructed from the updated QP eigenvalues using $\mathbf{k \cdot p}$ perturbation theory \cite{PAW_optics,JudithThesis2008,VASPguide}.
%
\subsection{\label{Optical Spectra Calculations}Optical Spectra Calculations}
Assuming that the orientation of the PdN(B24) crystal surface is parallel to the optical axis, all the frequency-dependent optical spectra (e. g.
 reflectivity $R\left(\omega\right)$,
 refractive index $n\left(\omega\right)$,
 energy-loss $L\left(\omega\right)$ and
 absorption coefficient $\alpha\left(\omega\right)$)
 can then straightforwardly be derived from the real $\varepsilon_{\text{re}}(\omega)$ and the imaginary $\varepsilon_{\text{im}}(\omega)$ parts of $\varepsilon_{\text{RPA}}(\omega)$ \cite{Fox,dressel2002electrodynamics,Ch9_in_Handbook_of_Optics_2010}: 
\begin{align}
R\left(\omega\right) &= \left| \frac{\left[  \varepsilon_{\text{re}}\left(\omega\right) + j \varepsilon_{\text{im}}\left(\omega\right)  \right]^{\frac{1}{2}} - 1}{\left[  \varepsilon_{\text{re}}\left(\omega\right) + j \varepsilon_{\text{im}}\left(\omega\right)  \right]^{\frac{1}{2}} + 1} \right| ^{2}		\label{R(omega)}\\
n\left(\omega\right) &= \frac{1}{\sqrt{2}} \left(  \left[  \varepsilon_{\text{re}}^{2}\left(\omega\right) + \varepsilon_{\text{im}}^{2}\left(\omega\right) \right]^{\frac{1}{2}} + \varepsilon_{\text{re}}\left(\omega\right) \right)^{\frac{1}{2}} 			\label{n(omega)}\\
L\left(\omega\right) &= \frac{\varepsilon_{\text{im}}\left(\omega\right)}{\varepsilon_{\text{re}}^{2}\left(\omega\right) + \varepsilon_{\text{im}}^{2}\left(\omega\right)} 			\label{L(omega)}\\
\alpha\left(\omega\right) &= \sqrt{2} \omega  \left(  \left[  \varepsilon_{\text{re}}^{2}\left(\omega\right) + \varepsilon_{\text{im}}^{2}\left(\omega\right) \right]^{\frac{1}{2}}  - \varepsilon_{\text{re}}\left(\omega\right) \right)^{\frac{1}{2}} 		\label{alpha(omega)}
\end{align}

We emphasized here that for one to obtain more accurate optical spectra (i.e., more accurate positions and amplitudes of the characteristic peaks), one should solve the equation of motion of the two-body Green's function $G_2$ (the so-called Bethe-Salpeter equation) in order to obtain the electron-hole excitations. $G_2$ can be evaluated on the basis of the $G_0W_0$-calculated one-particle Green's function $G$ and eigenvalues \cite{DFT_GW_BSE_Electron-hole_excitations_and_optical_spectra_from_first_principles_2000}.
%
\section{\label{Results and Discussion}Results and Discussion}
\begin{figure}
\includegraphics[width=0.45\textwidth]{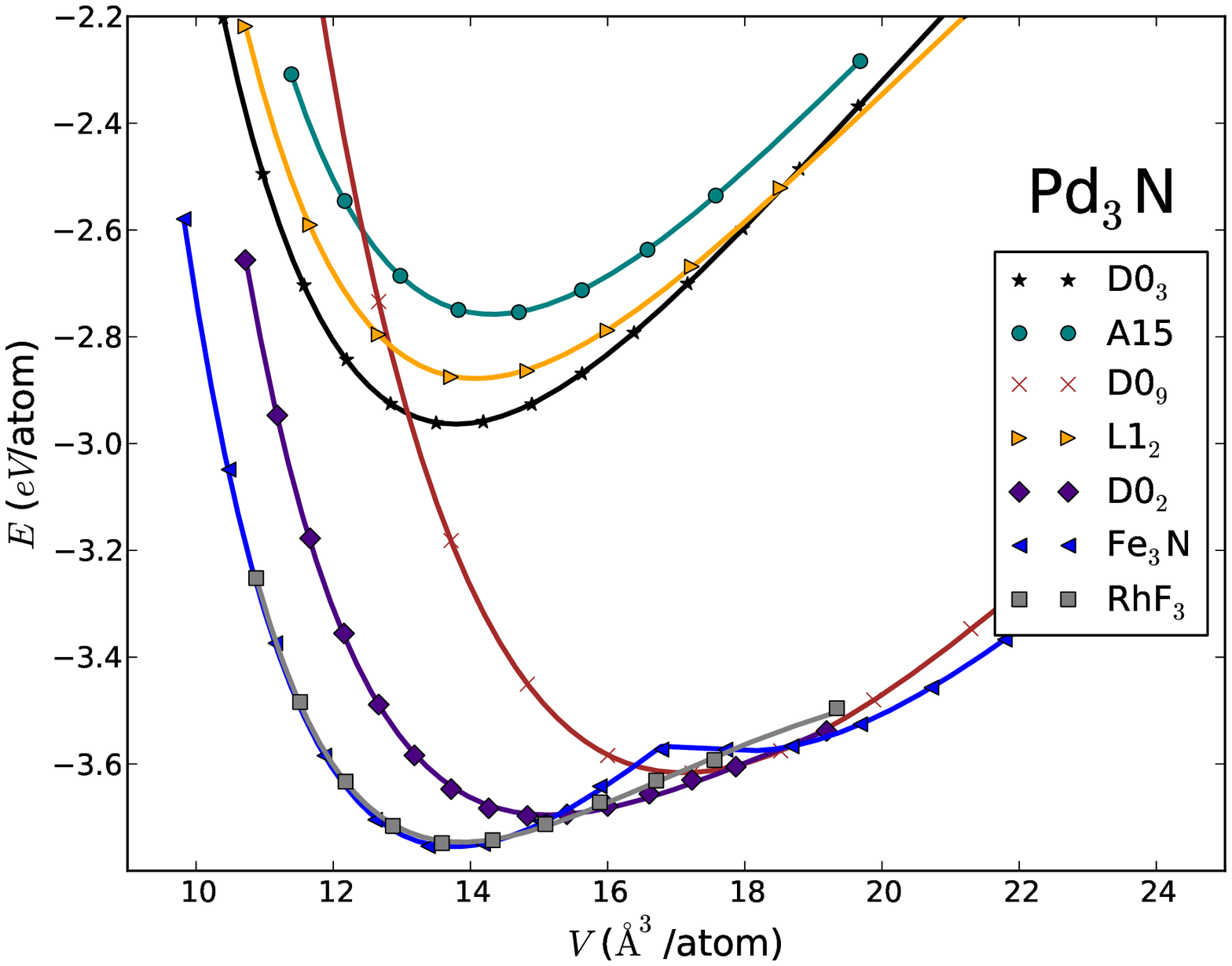}
\caption{\label{Pd3N1_ev_EOS}(Color online.) Cohesive energy $E_{\text{coh}} (eV/\text{atom})$ versus atomic volume $V$ (\text{\AA}$^{3}$/\text{atom}) for Pd$_3$N in seven different structural phases.}
\end{figure}

\begin{figure}
\includegraphics[width=0.45\textwidth]{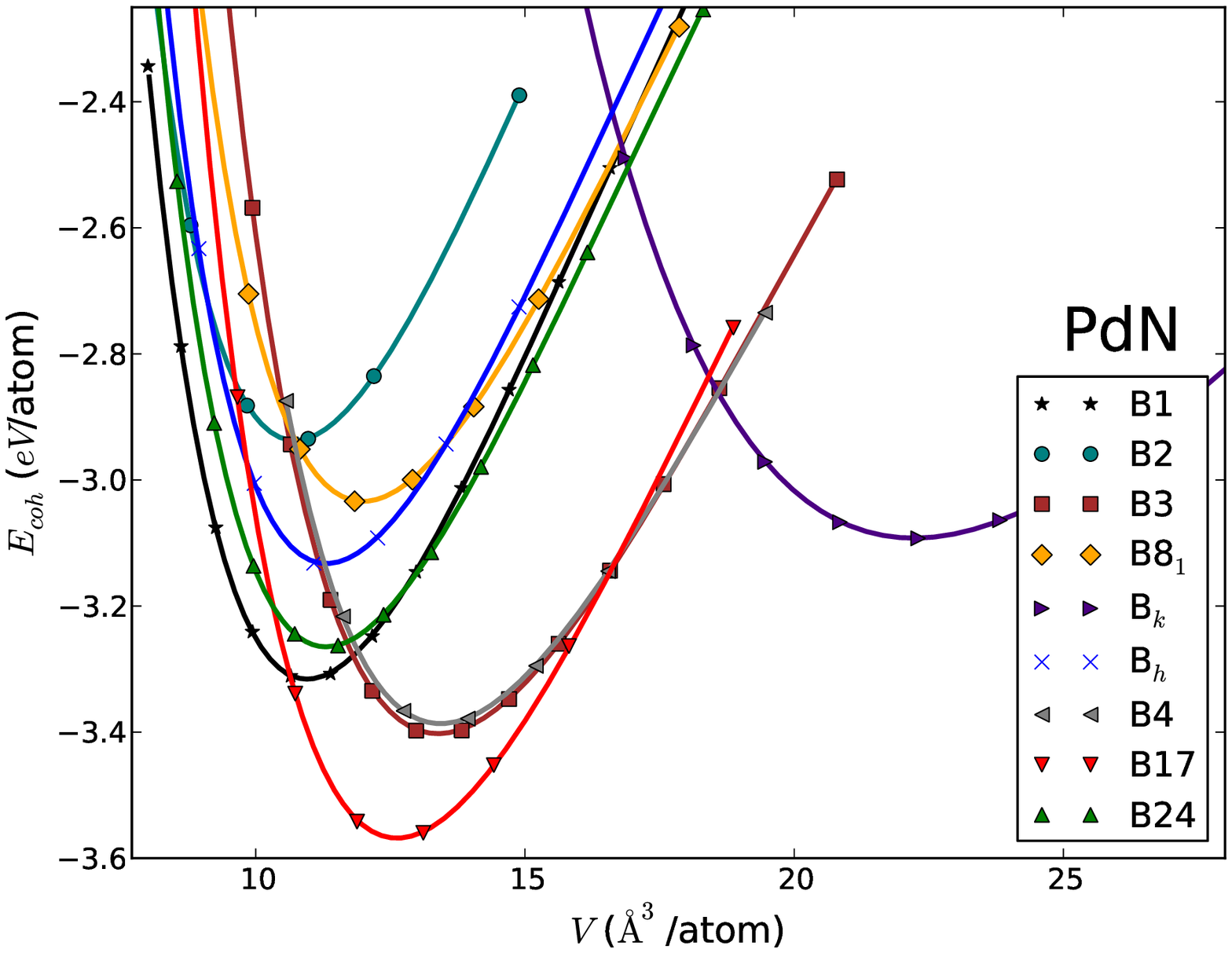}
\caption{\label{Pd1N1_ev_EOS}(Color online.) Cohesive energy $E_{\text{coh}} (eV/\text{atom})$ versus atomic volume $V$ (\text{\AA}$^{3}$/\text{atom}) for PdN in nine different structural phases.}
\end{figure}

\begin{figure}
\includegraphics[width=0.45\textwidth]{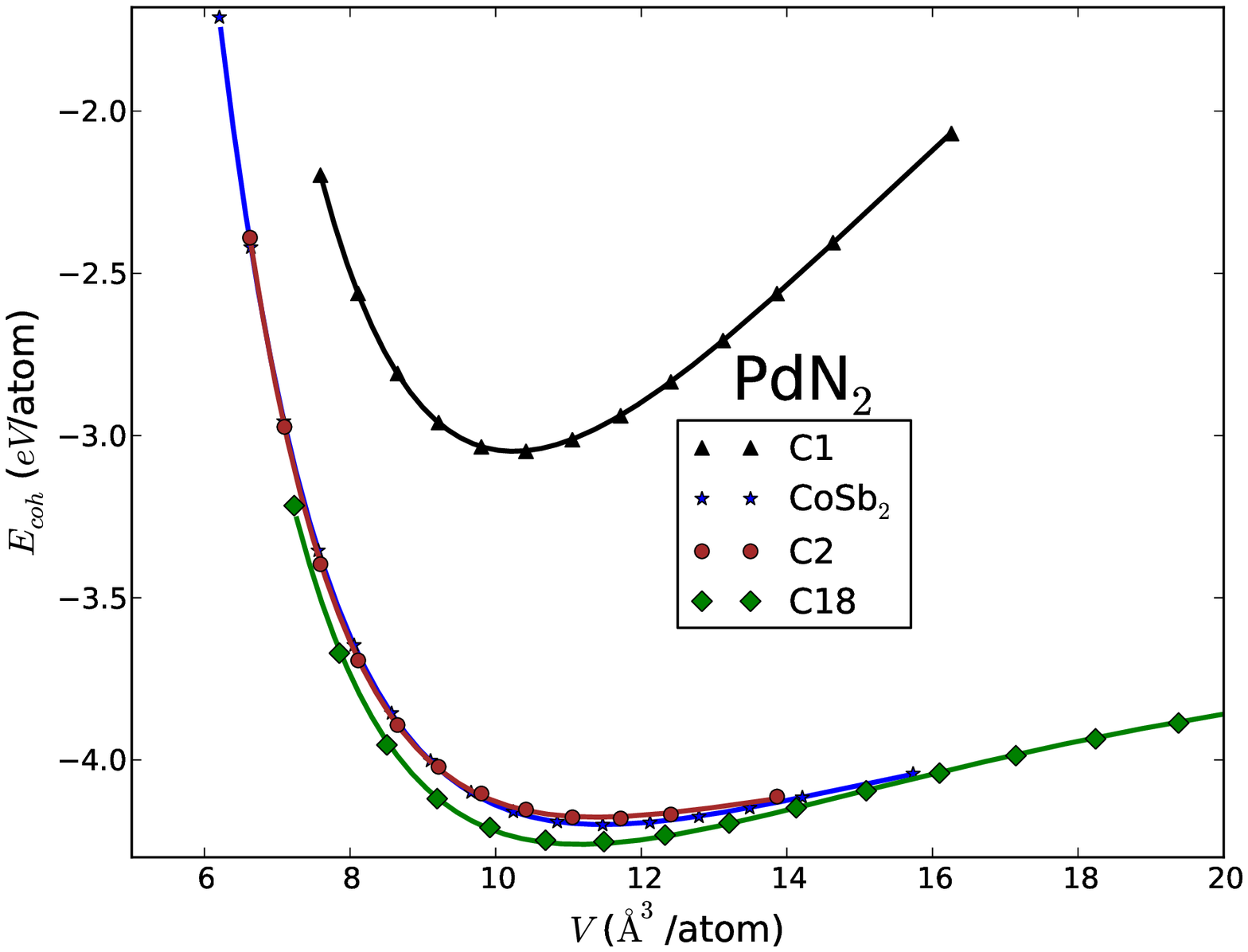}
\caption{\label{Pd1N2_ev_EOS}(Color online.) Cohesive energy $E_{\text{coh}} (eV/\text{atom})$ versus atomic volume $V (\text{\AA}^{3}/\text{atom})$ for PdN$_2$ in four different structural phases.}
\end{figure}

Cohesive energy $E_{\text{coh}}$ versus atomic volume $V_{0}$ equation of state (EOS) for the considered modifications of Pd$_3$N, PdN and PdN$_2$ are displayed graphically in Figs. \ref{Pd3N1_ev_EOS}, \ref{Pd1N1_ev_EOS} and \ref{Pd1N2_ev_EOS}, respectively. The corresponding calculated  equilibrium structural, energetic and mechanical properties of these twenty phases and of Pd(A1)are presented in Table \ref{table: palladium_nitrides_equilibrium_structural_properties}. Modifications in this table are ordered in the same way as in Table \ref{PdN's allstructures}. Our results are compared with experiment and with previous calculations. In the latter case, the calculations methods and the $XC$ functionals are indicated in the Table footnotes whenever possible.

\begin{table*}
\caption{\label{table: palladium_nitrides_equilibrium_structural_properties}Calculated and experimental zero-pressure properties of Pd(A1) and of the twenty studied phases of Pd$_{3}$N, PdN and PdN$_2$\textbf{:} Lattice constants ($a(\text{\AA})$, $b(\text{\AA})$, $c(\text{\AA})$, $\alpha(^{\circ})$ and $\beta(^{\circ})$), atomic volume $V_{0}(\text{\AA}^{3}/$atom$)$, cohesive energy $E_{\text{coh}} (eV/$atom$)$, bulk modulus $B_{0} (GPa)$ and its pressure derivative $B_{0}^{\prime}$, and formation energy $E_{\text{f}}(eV/\text{atom})$. The presented data are of the current work (\textit{Pres.}), experimentally reported (\textit{Expt.}) and of previous calculations (\textit{\textit{Comp.}}).}
\resizebox{1.0\textwidth}{!}{
\begin{tabular}{lllllllllll}
\hline		
\textbf{Structure}	&		& $a(\AA)$		   & $b(\AA)$			& $c(\AA)$		  & $\alpha(^{\circ})$ or $\beta(^{\circ})$		& $V_{0} (\AA^{3}/$atom$)$   & $E_{\text{coh}}(eV/\text{atom})$	& $B_{0}(\text{GPa})$		& $B_{0}^{\prime}$ & $E_{f}(eV/\text{atom})$\\
\hline \hline 
\multicolumn{11}{c}{\textbf{Pd}} \\
\hline 
\multirow{4}{*}{\textbf{A1}}       & \textit{Pres.}& $3.957$ & --    & --                       & --                        & $15.49$   & $-3.703$  & $163.626$ & $5.549$ & \\
        & \textit{Exp.} & $3.8900$\footnotemark[1] & --    & --                       & --                        & $14.716$\footnotemark[2]  & $-3.89$\footnotemark[3] & $180.8$\footnotemark[3], $184$\footnotemark[4]  & $5.42$\footnotemark[5]  & \\
        & \textit{Comp.} & $3.85$\footnotemark[6]\textsuperscript{,}\footnotemark[7] & --    & --                       & --    &  & $-5.06$\footnotemark[8],       & $212$\footnotemark[6], $220$\footnotemark[7] & $5.50$\footnotemark[9], $6.40$\footnotemark[10], & \\

        &    &      & --    & --      & --     & &   $-3.74$\footnotemark[11]\textsuperscript{,}\footnotemark[12]  &       &  $5.29$\footnotemark[13] & \\

\hline 
		\multicolumn{11}{c}{\textbf{Pd$_3$N}} \\
\hline 
\multirow{1}{*}{\textbf{D0$_3$}}        & \textit{Pres.}&  $6.043$    & --    & --                       & --                        & $13.79$  & $-2.965$  &  $162.895$   & $ 5.353$ & $1.111$ \\
	

\hline 
\multirow{1}{*}{\textbf{A15}}           & \textit{Pres.}& $4.857$ & --    & --                       & --                        &  $14.32$ &  $-2.758$  & $148.123$  & $5.321$ & $1.318$ \\
	

\hline 
\multirow{1}{*}{\textbf{D0$_9$}}        & \textit{Pres.}& $4.089$ & --    & --                       & --                        &   $17.09$  & $-3.617$ &  $132.499$ & $5.255$ & $0.459$ \\
	

\hline 
\multirow{1}{*}{\textbf{L1$_2$}}        & \textit{Pres.}& $3.834$ & --    & --                       & --                        &   $14.09$  &  $-2.880$  & $148.697$  & $5.402$ & $1.196$ \\
	

\hline 
\multirow{1}{*}{\textbf{D0$_2$}}        & \textit{Pres.}& $7.828$ & --    & --                       & --                        &  $14.99$ & $-3.698$  &   $111.276$    & $9.361$ & $0.378$ \\
	

\hline 
\multirow{1}{*}{\textbf{$\epsilon$-Fe$_3$N}}& \textit{Pres.}& $5.135$ & --    & $4.785$     & --     				&  $13.66$    & $-3.758$  &  $168.259$   & $8.694$ & $0.318$ \\
	

\hline 
\multirow{1}{*}{\textbf{RhF$_3$}}       & \textit{Pres.}& $5.627$ & --    & --        & $\alpha = 54.640$                        & $13.78$  &  $-3.749$   &	$130.415$	& $9.837$ & $0.327$ \\
	

\hline 
					\multicolumn{11}{c}{\textbf{PdN}} \\
\hline 
\multirow{4}{*}{\textbf{B1}}   		& \textit{Pres.}&  $4.444$  & --    & --                       & --                        &  $10.97$  &  $-3.317$  &   $207.787$	& $4.978$ & $1.132$ \\
	
        & \textit{Comp.} & $4.145$\footnotemark[14]  & --    & --                       & --                        &                           & $-4.585$\footnotemark[15]           &        &   & $0.400$\footnotemark[17]\\

        & 		 & 	$4.33$\footnotemark[18]  & --    & --                       & --                        &                           &  $-11.90$\footnotemark[18]   &  $297.67$\footnotemark[18]      & $4.15$\footnotemark[18]  & 		\\

        & 		 & 				  & --    & --                       & --                        &                           & $-4.027$\footnotemark[16]           &        &   & 		\\

\hline 
\multirow{2}{*}{\textbf{B2}}   		& \textit{Pres.}& $2.779$ & --    & --                       & --                        &   $10.73$ &   $-2.947$  & $210.200$ & $4.931$ & $1.502$ \\

        & \textit{Comp.} & 	$2.71$\footnotemark[18]  & --    & --                       & --                        &                           &  $-12.25$\footnotemark[18]   &  $251.03$\footnotemark[18]      & $4.70$\footnotemark[18]  & 		\\

\hline 
\multirow{2}{*}{\textbf{B3}}   		& \textit{Pres.}& $4.748$ & --    & --                       & --                        &  $13.38$  &  $-3.404$ &  $167.804$   & $5.015$ & $1.045$ \\
	
        & \textit{Comp.} & 	$4.67$\footnotemark[18]  & --    & --                       & --                        &                           &  $-8.89$\footnotemark[18]   &  $192.33$\footnotemark[18]      & $4.07$\footnotemark[18]  & 		\\

\hline 
\multirow{1}{*}{\textbf{B8$_{1}$}}   	& \textit{Pres.}	&  $3.416$ & --    & $4.751$   & --                        & $12.00$ &  $-3.034$  & $187.954$  & $5.021$ & $1.415$ \\
	

\hline 
\multirow{1}{*}{\textbf{B$_{\text{k}}$}}& \textit{Pres.}	& $3.378$ & --      & $8.986$     & --                   & $22.20$ &  $-3.092$  &  $88.897$ & $4.830$ & $1.357$ \\
	

\hline 
\multirow{1}{*}{\textbf{B$_{\text{h}}$}}& \textit{Pres.}	& $2.992$ & --    & $2.921$       & --    				&  $11.32$ &  $-3.135$   & $201.682$ & $5.037$ & $1.314$\\
	

\hline 
\multirow{2}{*}{\textbf{B4}}   		& \textit{Pres.}	& $3.360$   & --    & $5.503$     & --                        &  $13.45$  & $-3.387$ &  $164.169$ & $4.978$ & $1.062$ \\

        & \textit{Comp.} & 	$3.37$\footnotemark[18]  & --    & $5.26$\footnotemark[18]  & --                        &                           &  $-11.43$\footnotemark[18]   &  $171.34$\footnotemark[18]      & $4.63$\footnotemark[18]  & 		\\

\hline 
\multirow{1}{*}{\textbf{B17}}   	& \textit{Pres.}	& $3.061$ & --    & $5.389$      & --                        &   $12.62$   & $-3.570$ & $190.426$ & $4.993$ & $0.879$ \\


\hline 
\multirow{1}{*}{\textbf{B24}}   	& \textit{Pres.}	& $4.173$  & $4.427$    & $4.898$  & --                        &  $11.31$  &  $-3.265$  & $197.566$ & $4.997$ & $1.184$ \\

\hline 
					\multicolumn{11}{c}{\textbf{PdN$_2$}} \\
\hline 
\multirow{1}{*}{\textbf{C1}}            & \textit{Pres.}	& $4.975$ & --    & --                       & --                        &  $10.26$ &  $-3.050$ & $221.734$  & $4.809$ & $1.648$ \\


\hline 
\multirow{3}{*}{\textbf{C2}}            & \textit{Pres.}	& $5.169$ & --    & --                       & --                        &   $11.51$  & $-4.181$  & $68.462$ & $5.611$ & $0.517$ \\


        & \textit{Comp.} & $4.975$\footnotemark[19]	&--	&--	&--	&$10.267$\footnotemark[19]	&--	&$135$\footnotemark[19]	&		& \\
        & 				 & $4.843$\footnotemark[20]	&--	&--	&--	&			&--	&$156$\footnotemark[20]	&$9.48$\footnotemark[20]	&	\\

\hline 
\multirow{2}{*}{\textbf{C18}}           & \textit{Pres.}	& $3.173$ & $4.164$   & $5.082$   & --                        &   $11.19$  &  $-4.254$  & $76.615$ & $6.102$ & $0.444$ \\

        & \textit{Comp.} & $3.911$\footnotemark[19]	& $4.975$\footnotemark[19]	& $3.133$\footnotemark[19]	&--	&$10.333$\footnotemark[19]	&--	&$100$\footnotemark[19]	&		& \\

\hline 
\multirow{2}{*}{\textbf{CoSb$_2$}}      & \textit{Pres.}	& $5.608$ & $5.304$    & $9.630$  & $\beta = 151.225$                        &  $11.49$  &   $-4.200$   & $71.792$ & $6.511$ & $0.498$ \\

        & \textit{Comp.} & $5.071$\footnotemark[19]	& $5.005$\footnotemark[19]	& $5.071$\footnotemark[19]	&--	&$10.433$\footnotemark[19]	&    	&$93$\footnotemark[19]	&		& \\

\hline \hline 
\end{tabular}
}       

\footnotetext[1]{\;Ref. \onlinecite{Jerry_1974}: This is an average of 21 experimental values, at $20 ^{\circ} C$, with a deviation $\pm 0.0007$ \AA.}
\footnotetext[2]{\;Ref. \onlinecite{Jerry_1974}: At $20 ^{\circ} C$.}
\footnotetext[3]{\;Ref. \onlinecite{Kittel}: Cohesive energies are given at $0 \; K$ and $1 \text{ atm} = 0.00010 \text{GPa}$; while bulk moduli are given at room temperature.}
\footnotetext[4]{\;Ref. (25) in \onlinecite{B_prime_1997_theory_comp_n_exp}: at room temperature.}
\footnotetext[5]{\;See Refs. (8)--(11) in \onlinecite{B_prime_1997_theory_comp_n_exp}.}
\footnotetext[6]{\;Ref. \onlinecite{elemental_metals_1996_comp}. LAPW-TB.}
\footnotetext[7]{Ref. \onlinecite{elemental_metals_1996_comp}. LAPW-LDA.}
\footnotetext[8]{\;Ref. \onlinecite{elemental_metals_2008_comp}: PAW-LDA.}
\footnotetext[9]{\;Ref. \onlinecite{B_prime_1997_theory_comp_n_exp}: Using the so-called method of transition metal pseudopotential theory; a modified form of a method proposed by Wills and Harrison to represent the effective interatomic interaction.}
\footnotetext[10]{\;Ref. \onlinecite{B_prime_1997_theory_comp_n_exp}: Using a semi-empirical estimate based on the calculation of the slope of the shock velocity \textit{vs.} particle velocity curves obtained from the dynamic high-pressure experiments. The given values are estimated at $\sim 298 \; K$.}
\footnotetext[11]{\;Ref. \onlinecite{elemental_metals_2008_comp}: PAW-PW91.}
\footnotetext[12]{\;Ref. \onlinecite{elemental_metals_2008_comp}: PAW-PBE.}
\footnotetext[13]{\;Ref. \onlinecite{B_prime_1997_theory_comp_n_exp}: Using a semi-empirical method in which the experimental static $P-V$ data are fitted to an EOS form where $B_{0}$ and $B_{0}^{\prime}$ are adjustable parameters.  The given values are estimated at $\sim 298 \; K$.}

\footnotetext[14]{\;Ref. \onlinecite{PdN_1992_exp_n_comp}: Estimated by extrapolation of the (experimental) average volume per atom $\Omega_{\text{MN}}$ for nitrides of other $4d$ transition metals.}
\footnotetext[15]{\;Ref. \onlinecite{PdN_1992_exp_n_comp}: Using the linear-muffin-tin-orbitals (LMTO) method and the local-spin-density approximation (LSDA).}
\footnotetext[16]{\;Ref. \onlinecite{PdN_1992_exp_n_comp}: ($ \pm 0.150$) Semi-empirical calculations.} 
\footnotetext[17]{\;Ref. \onlinecite{PdN_1992_exp_n_comp}: This is enthalpy of formation ($ \pm 0.145$) from Pd and N in their stable modifications at one atmosphere and $T = -298.15 \; K$.}

\footnotetext[18]{\;Ref. \onlinecite{PdN_2010_comp}: Using separable norm-conserving pseudopotentials within LDA.}

\footnotetext[19]{\;Ref. \onlinecite{PdN2_2010_comp_5+}: Using the Vanderbilt ultrasoft pseudopotential within GGA.}

\footnotetext[20]{\;Ref. \onlinecite{PdN2_n_PtN2_2009_comp}: Using PAW within LDA.}

\end{table*}

\begin{figure*}
\includegraphics[width=0.95\textwidth]{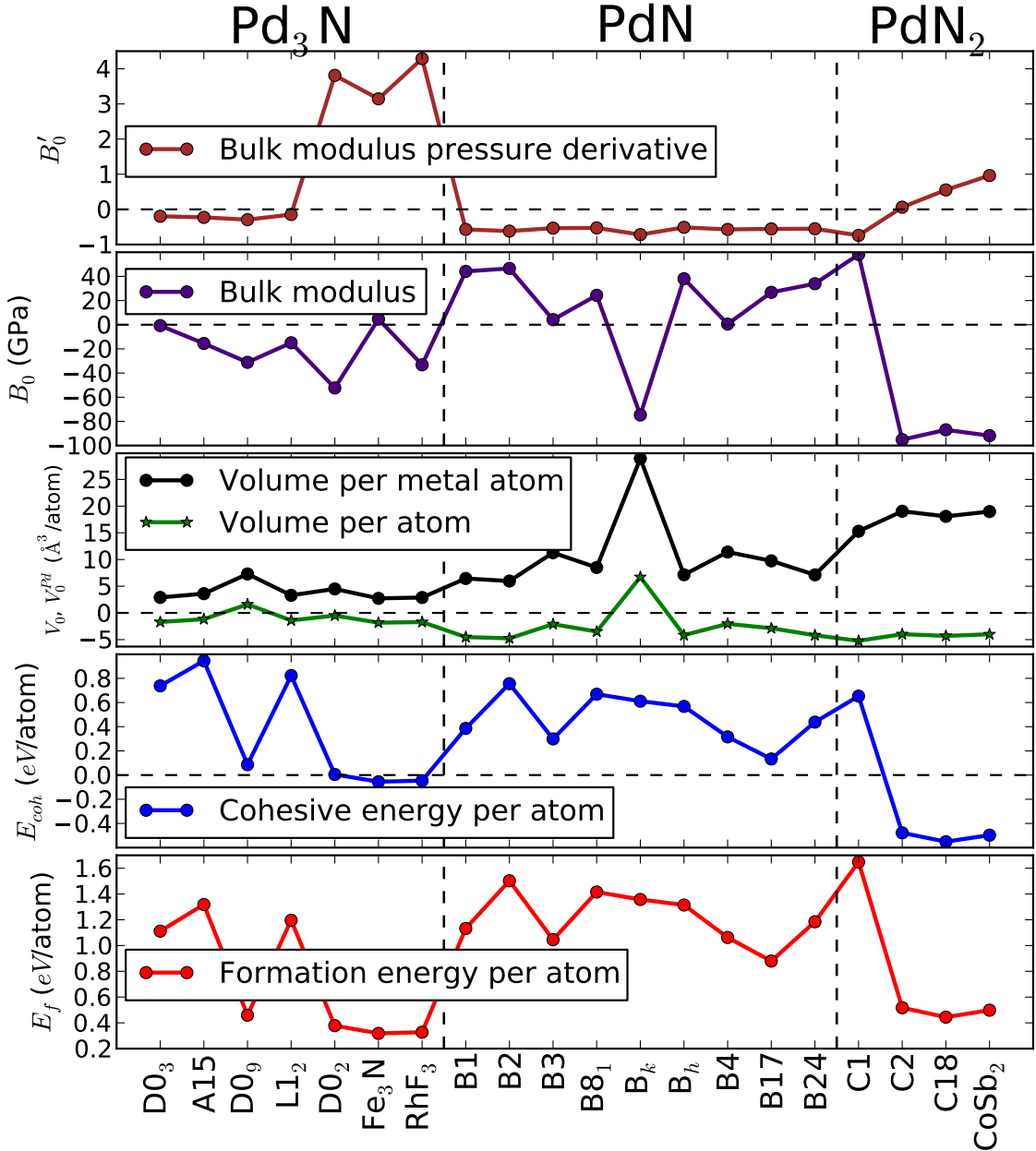}
\caption{\label{figure: palladium_nitrides_equilibrium_properties}(Color online.) Calculated equilibrium properties of the twenty studied phases of palladium nitrides. All quantities are given relative to the corresponding ones of the \textit{fcc} crystalline elemental palladium given in the first row of Table \ref{table: palladium_nitrides_equilibrium_structural_properties}. The vertical dashed lines separate between the different stoichoimetries.}
\end{figure*}

To compare and to deeper analyze the obtained equilibrium properties of the three stoichiometries series with respect to one another, the calculated equilibrium properties are depicted graphically in Fig. \ref{figure: palladium_nitrides_equilibrium_properties}. All quantities in this figure are given relative to the corresponding ones of Pd(A1) given in Table \ref{table: palladium_nitrides_equilibrium_structural_properties}. In this way, one will be able to investigate the effect of nitridation on the parent crystalline Pd as well \footnote{In Table \ref{table: palladium_nitrides_equilibrium_structural_properties}, our computed properties of the elemental Pd are compared with experiment and with previous calculations as well. This may benchmark the accuracy of the rest of our calculations.}.

%
\subsection{\label{EOS and Relative Stabilities}EOS and Relative Stabilities}
In Fig. \ref{Pd3N1_ev_EOS}, the energy-volume EOSs of the seven considered Pd$_3$N modifications are displayed. This figure, and the values of the equilibrium cohesive energy $E_{\text{coh}}$ (Table \ref{table: palladium_nitrides_equilibrium_structural_properties} and Fig. \ref{figure: palladium_nitrides_equilibrium_properties}) reveal that the Fe$_3$N structure (of the Ni$_3$N) is the most energetically favored modification, as we expected. However, the rhombohedric RhF$_3$ phase has a very similar EOS curve before and around the equilibrium, with very close $E_{\text{coh}}$ value to that of Fe$_3$N. Cubic systems (D0$_3$, A15, D0$_9$, L1$_2$ and D0$_2$) seem not to be energetically competing in this stoichiometry.

It is clear that the simple tetragonal structure of cooperite (B17) would be the energetically most stable phase of PdN (Fig. \ref{Pd1N1_ev_EOS}). To the best of our knowledge, this structure has not been considered for PdN in any earlier work, though it was theoretically predicted to be the ground-state structure of the nitrides of the elements surrounding Pd in the periodic table: 
 PtN \cite{Suleiman_PhD_arxiv_copper_nitrides_article,Mysterious_Platinum_Nitride_2006_comp},
 CuN \cite{Suleiman_PhD_arxiv_copper_nitrides_article},
 AgN \cite{Suleiman_PhD_arxiv_silver_nitrides_article},
 and AuN \cite{Suleiman_PhD_SAIP2012_gold_nitrides_article}.
Nevertheless, Fig. \ref{figure: palladium_nitrides_equilibrium_properties} shows clearly that no PdN phase, even PdN(B17), has a tendency to lower the cohesive energy of the parent metal.
 
In Ref. \onlinecite{PdN_2010_comp} the $E(V)$ EOS for PdN in the B1, B2, B3 and B4 structures was studied. Within this parameter sub-space, the relative stabilities arrived at in that work agree very well with ours. However, their obtained $E_{\text{coh}}$ are more than twice the values we obtained, and the bulk moduli differ considerably (see Table \ref{table: palladium_nitrides_equilibrium_structural_properties})!

In the studied parameter sub-space of PdN$_{2}$, the marcasite structure (C18) is the most energetically stable. The relative stability of C2 and CoSb$_{2}$ phases may be compared with Crowhurst \textit{et al.} \cite{PdN2_2008_exp_n_comp} who found PdN$_{2}$ in the baddeleyite structure (which is very close to CoSb$_{2}$ structure \cite{Synthesis_of_Binary_TMNs_2011_exp_review_5+}) to be more stable than PdN$_{2}$(C2).

From a combined theoretical and experimental investigation, \AA{}berg \textit{et al.} \cite{PdN2_2010_exp_n_comp} showed that for PdN$_{2}$(C2) both the electronic and the structural degrees of freedom have a strong pressure dependence. They claimed that the EOS \textit{cannot} be accurately described within the GGA. Earlier calculations showed that PdN$_{2}$(C2) is very soft (see Ref. 22 in \onlinecite{PdN2_2008_exp_n_comp}). These two facts may explain the difficulty we found in relaxing this structure as well as they may explain the considerable differences found with and among the earlier reported structural properties.
%
\subsection{\label{Volume per Atom and Lattice Parameters}Volume per Atom and Lattice Parameters}
From Fig. \ref{figure: palladium_nitrides_equilibrium_properties} one can see clearly that except Pd$_3$N(D0$_9$) and PdN(B$_k$), all phases tend to lower the \textit{volume per atom} of their parent metal. The metal-metal bond length, as represented by the volume per \textit{metal} atom $V_0^{\text{Pd}}$, increases (on average) in the direction of increasing nitrogen content and decreasing structural symmetry.
%
\subsection{\label{Pressure-Induced Phase Transitions}Pressure-Induced Phase Transitions}
\begin{figure}
\includegraphics[width=0.45\textwidth]{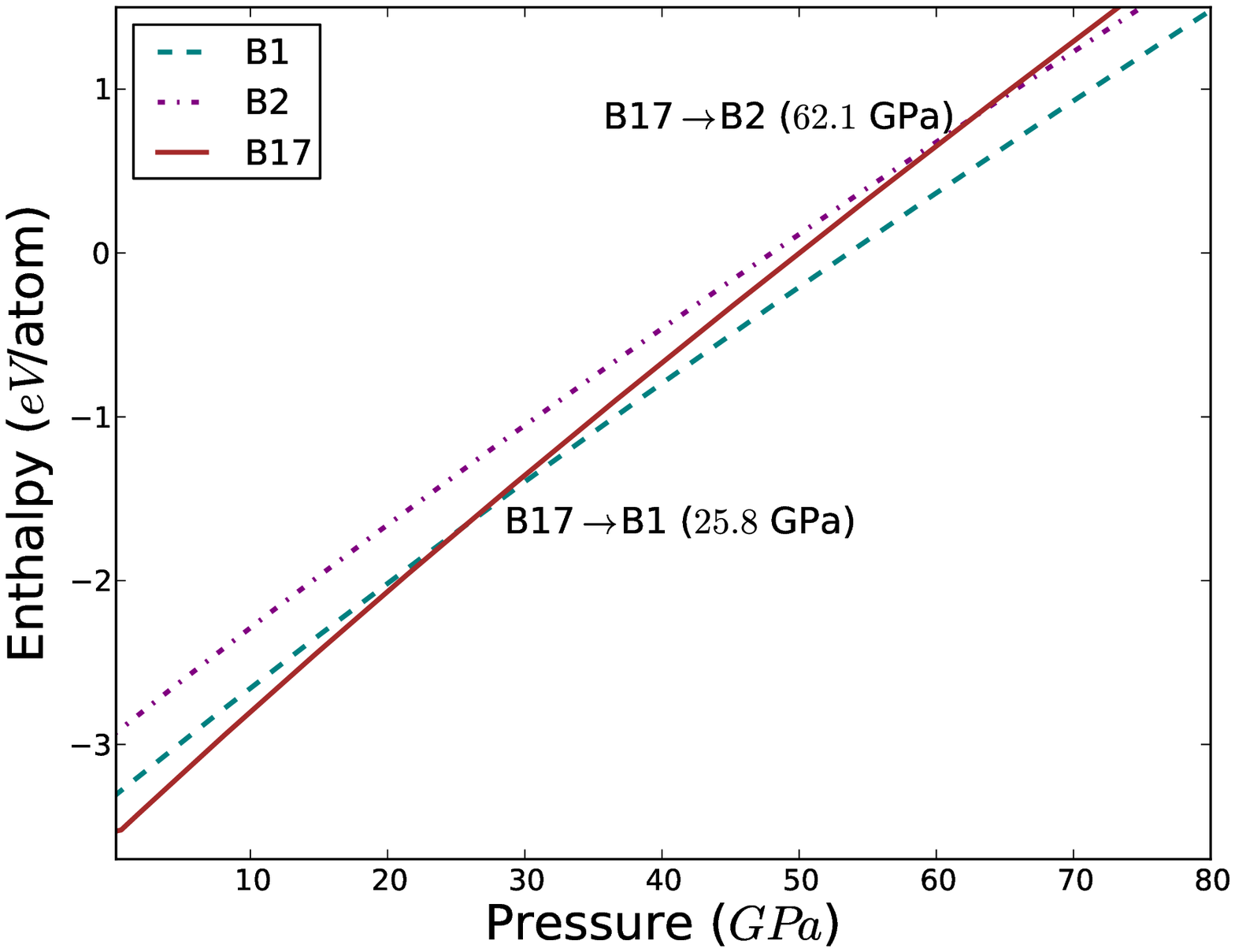}
\includegraphics[width=0.45\textwidth]{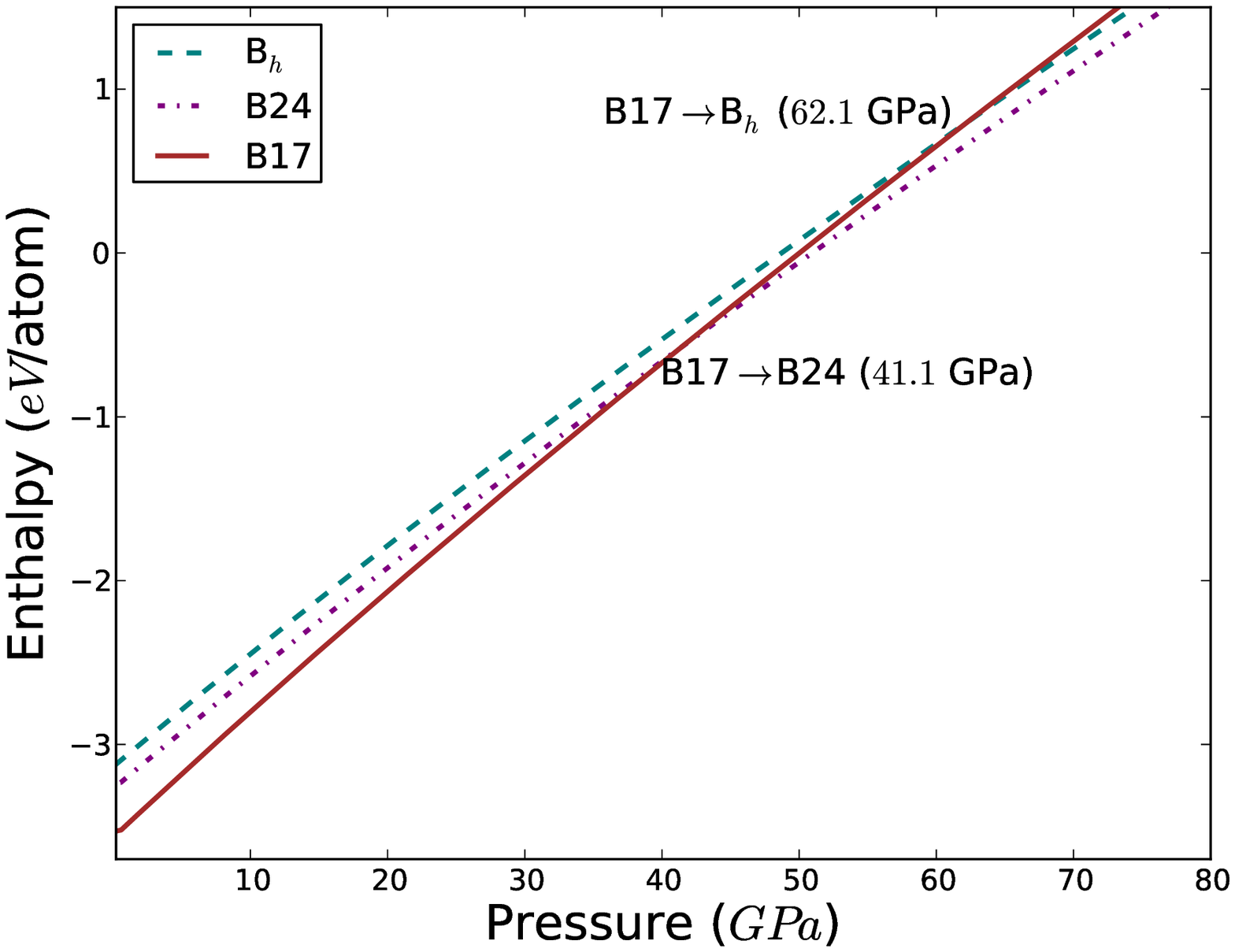}
\caption{\label{Pd1N1_phase_transitions}(Color online.) Enthalpy \textit{vs.} pressure for some PdN modifications in the region where: B17$\rightarrow$B1 and B17$\rightarrow$B2 (top) and B17$\rightarrow$B$_{h}$ and B17$\rightarrow$B24 (bottom) phase transitions occur.}
\end{figure}

Enthalpy-pressure relations for PdN in some of the considered structures are displayed in Figs. \ref{Pd1N1_phase_transitions}. A point at which enthalpies $H = E_{\text{coh}}(V ) + PV$ of two structures are equal defines the transition pressure $P_{t}$, where transition from the phase with higher enthalpy to the phase with lower enthalpy may occur.

Some possible transitions and the corresponding $P_{t}$'s are depicted in Fig. \ref{Pd1N1_phase_transitions}. From the top subfigure, it is clear that, in this parameter sub-space, B17 structure is preferred at pressures below $\sim 25$ GPa, while B1 structure, the most popular structure for transition-metal mono-nitrides, is favoured at higher pressures. The bottom subfigure reveals that B24 is favored over B17 and B$_h$ at pressures higher than $41$ GPa.


%
\subsection{\label{Bulk Modulus and its Pressure Derivative}Bulk Modulus and its Pressure Derivative}
Fig. \ref{figure: palladium_nitrides_equilibrium_properties} shows that the bulk moduli of the Pd$_3$N phases, except Pd$_3$N(Fe$_3$N), tend to be lower than that of Pd, while 1:1 nitrides, except (B$_k$) tend to increase it. Despite the lower $V_0$ and the lower $E_{\text{coh}}$ possessed by the last three PdN$_2$ phases, they have $\sim 100 \; \mathrm{GPa}$ lower $B_0$ than their parent metal. This can be understood only in terms of the increase in the metal-metal bond length (represented by $V_0^{\text{Pd}}$).

Upon application of external pressure, the first four Pd$_3$N phases, all PdN phases and PdN$_2$(C1) phase tend to lower their $B_0$. PdN$_2$(C2) has the same sensitivity of its parent metal. PdN$_2$(C18 and CoSb$_2$) tend to increase their $B_0$. Pd$_3$N(D0$_2$, Fe$_3$N and RhF$_3$), however, are far more sensitive to external pressure, and their bulk moduli tend to increase significantly under pressure.
%
\subsection{\label{Thermodynamic Stability}Thermodynamic Stability}
Interestingly, Fig. \ref{figure: palladium_nitrides_equilibrium_properties} reveals that Pd$_3$N(Fe$_3$N and RhF$_3$) phases have lower formation energy than all PdN and PdN$_2$. That is, Pd$_3$N(Fe$_3$N and RhF$_3$) are thermodynamically favored over PdN and PdN$_2$. Nevertheless, the PdN$_2$ modifications, except C1, have significantly lower cohesive energy than the most favored Pd$_3$N phases. The numerical values of the formation energy (Table \ref{table: palladium_nitrides_equilibrium_structural_properties}) and and their graphical representation (Fig. \ref{figure: palladium_nitrides_equilibrium_properties}) reveal that it may be relatively harder to form a 1:1 palladium nitride.
%
\subsection{\label{Electronic Properties}Electronic Properties}
\begin{figure*}
\includegraphics[width=1.0\textwidth]{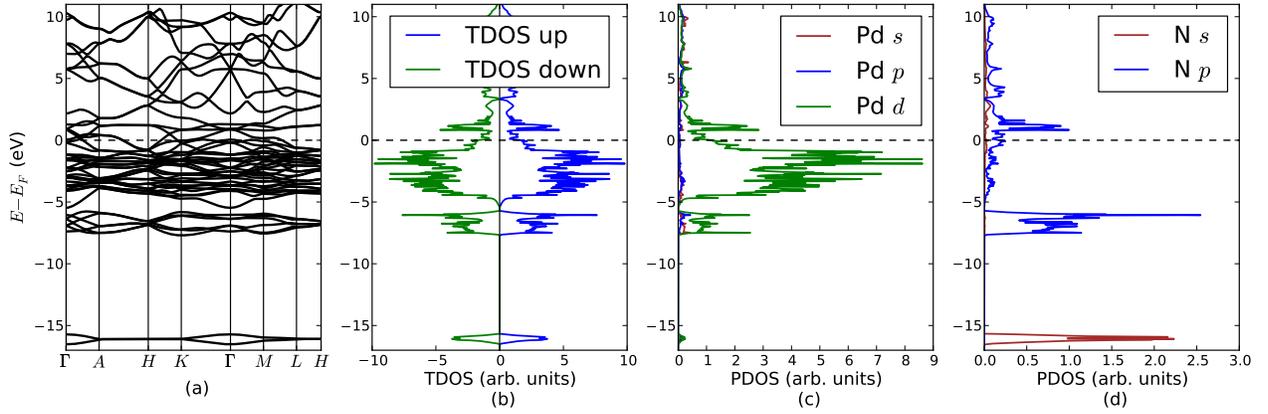}
\caption{\label{Pd3N1_Fe3N_electronic_structure}(Color online.) DFT calculated electronic structure for Pd$_3$N in the Fe$_3$N structure: \textbf{(a)} band structure along the high-symmetry $\mathbf{k}$-points which are labeled according to Ref. [\onlinecite{Bradley}]. Their coordinates w.r.t. the reciprocal lattice basis vectors are: $\Gamma(0, 0, 0)$, $A(0, 0, \frac{1}{2})$, $H(-\frac{1}{3} , \frac{2}{3} , \frac{1}{2})$ $K(-\frac{1}{3} ,\frac{2}{3} , 0)$, $M(0, \frac{1}{2}, 0)$, $L(0, \frac{1}{2}, \frac{1}{2})$, 
; \textbf{(b)} spin-projected total density of states (TDOS); \textbf{(c)} partial density of states (PDOS) of Pd($s, p, d$) orbitals in Pd$_3$N; and \textbf{(d)} PDOS of N($s, p$) orbitals in Pd$_3$N.}
\end{figure*}

\begin{figure*}
\includegraphics[width=1.0\textwidth]{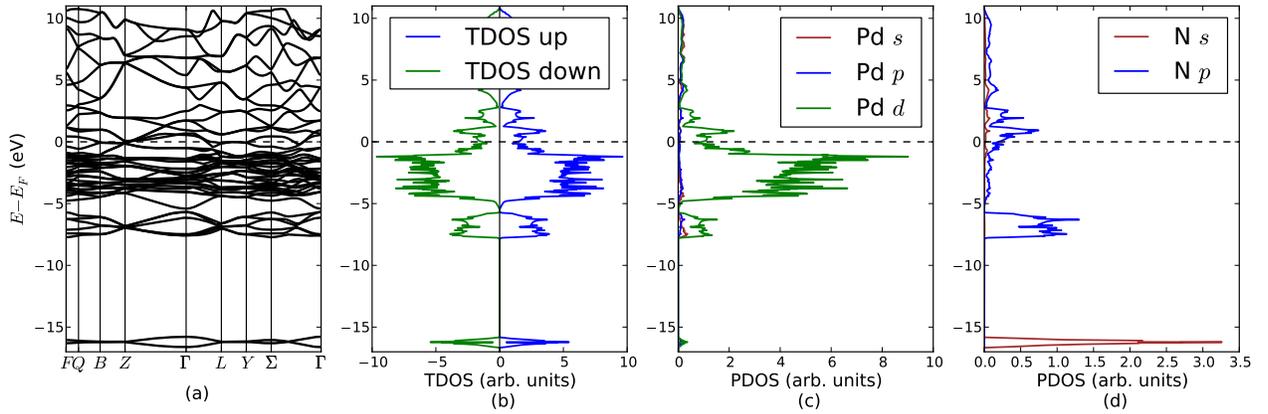}
\caption{\label{Pd3N1_RhF_3_electronic_structure}(Color online.) DFT calculated electronic structure for Pd$_3$N in the RhF$_3$ structure: \textbf{(a)} band structure along the high-symmetry $\mathbf{k}$-points which are labeled according to Ref. [\onlinecite{Bradley}]. Their coordinates w.r.t. the reciprocal lattice basis vectors are: $F(0.5, 0.5, 0.0)$, $Q(0.375, 0.625, 0.0)$, $B(0.5, 0.75, 0.25)$, $Z(0.5, 0.5, 0.5)$, $\Gamma(0.0,  0.0, 0.0)$, $L(0.0, 0.5, 0.0)$, $Y(0.25, 0.5, -.25)$, $\Sigma(0.0, 0.5, -.5)$; \textbf{(b)} spin-projected total density of states (TDOS); \textbf{(c)} partial density of states (PDOS) of Pd($s, p, d$) orbitals in Pd$_3$N;
 and \textbf{(d)} PDOS of N($s, p$) orbitals in Pd$_3$N.}
\end{figure*}

With the Fermi surface crossing the partly occupied bands, it is evident from Figs. \ref{Pd3N1_RhF_3_electronic_structure} and \ref{Pd3N1_Fe3N_electronic_structure} that Pd$_3$N(Fe$_3$N and RhF$_3$) are metals. In both cases, the strong Pd($d$)-N($p$) mixture is in the range ($-7.7 \sim -5.7 \; eV$). The Pd($d$)-N($p$) hybridization in the range ($-5 \sim E_F \; eV$) has very little contribution from the N($p$) states.

\begin{figure*}
\includegraphics[width=1.0\textwidth]{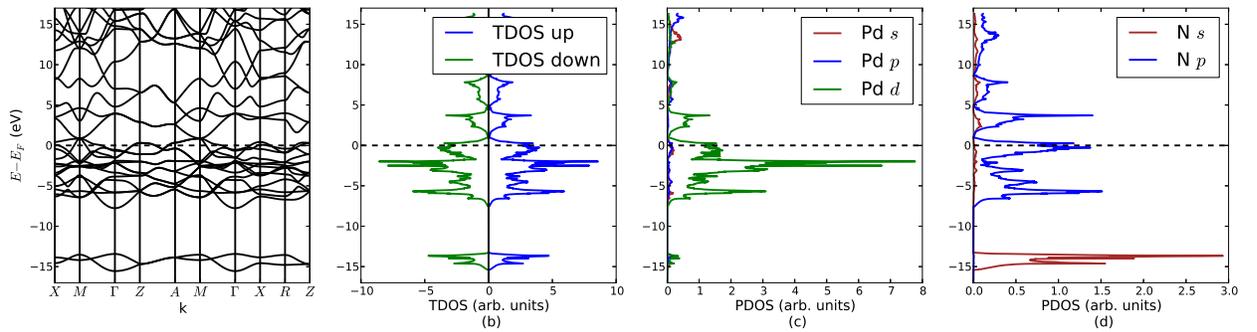}
\caption{\label{Pd1N1_B17_electronic_structure}(Color online.) DFT calculated electronic structure for PdN in the B17 structure:
 \textbf{(a)} band structure along the high-symmetry $\mathbf{k}$-points which are labeled according to Ref. [\onlinecite{Bradley}]. Their coordinates w.r.t. the reciprocal lattice basis vectors are: $X (0.0, 0.5, 0.0)$, $M (0.5, 0.5, 0.0)$, $\Gamma (0.0, 0.0, 0.0)$, $Z (0.0, 0.0, 0.5)$, $A (0.5, 0.5, 0.5)$, $R (0.0, 0.5, 0.5)$;
 \textbf{(b)} spin-projected total density of states (TDOS);
 \textbf{(c)} partial density of states (PDOS) of Pd$s, p, d$) orbitals in PdN;
 and \textbf{(d)} PDOS of N($s, p$) orbitals in PdN.}
\end{figure*}

\begin{figure*}
\includegraphics[width=1.0\textwidth]{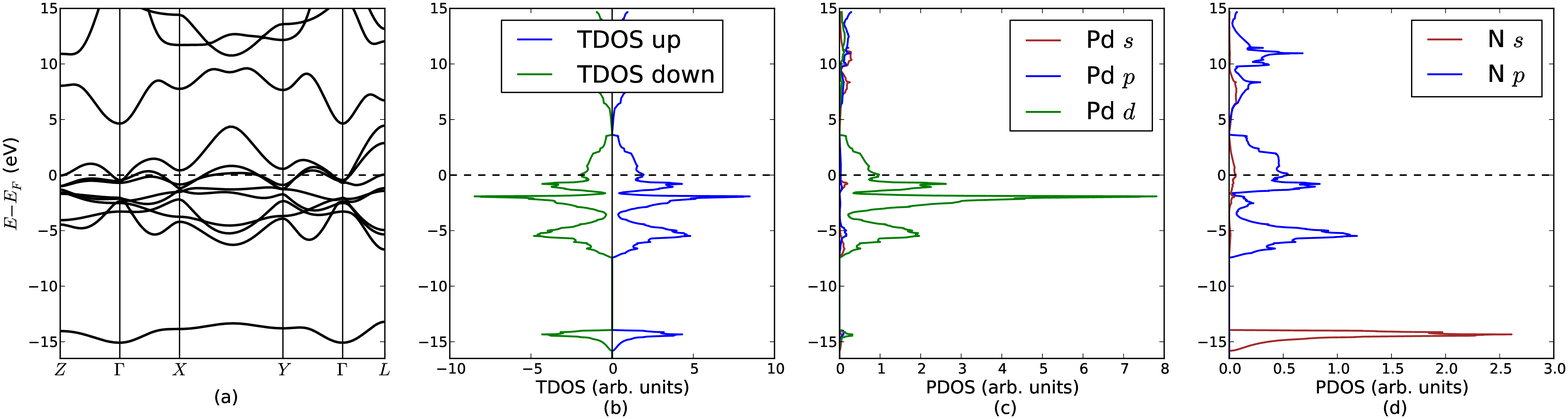}
\caption{\label{Pd1N1_B24_electronic_structure}(Color online.) DFT calculated electronic structure for PdN in the B24 structure:
 \textbf{(a)} band structure along the high-symmetry $\mathbf{k}$-points which are labeled according to Ref. [\onlinecite{Bradley}]. Their coordinates w.r.t. the reciprocal lattice basis vectors are: $Z (0.5, 0.5, 0.0)$, $\Gamma (0.0, 0.0, 0.0)$, $X (0.5, 0.0, 0.5)$, $Y (0.0, -.5, -.5)$ and $L (0.5, 0.0, 0.0)$; \textbf{(b)} spin-projected total density of states (TDOS);
 \textbf{(c)} partial density of states (PDOS) of Pd($s, p, d$) orbitals in PdN;
 and \textbf{(d)} PDOS of N($s, p$) orbitals in PdN.}
\end{figure*}

\begin{figure*}
\includegraphics[width=1.0\textwidth]{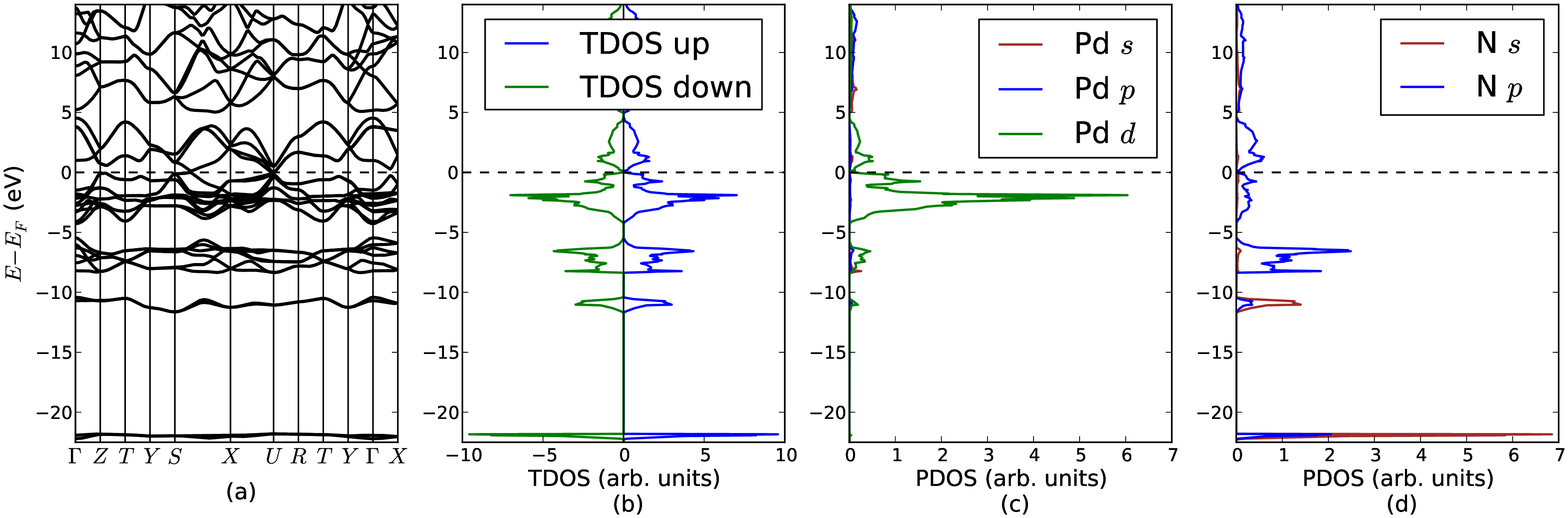}
\caption{\label{Pd1N2_C18_electronic_structure}(Color online.) DFT calculated electronic structure for PdN$_2$ in the C18 structure:
 \textbf{(a)} band structure along the high-symmetry $\mathbf{k}$-points which are labeled according to Ref. [\onlinecite{Bradley}]. Their coordinates w.r.t. the reciprocal lattice basis vectors are: $\Gamma( 0.0, 0.0, 0.0)$, $X( 0.0, 0.5, 0.0)$, $S( -.5, 0.5, 0.0)$, $Y( -.5, 0.0, 0.0)$, $Z( 0.0, 0.0, 0.5)$, $U( 0.0, 0.5, 0.5)$, $R( -.5, 0.5, 0.5)$, $T( -.5, 0.0, 0.5)$;
 \textbf{(b)} spin-projected total density of states (TDOS);
 \textbf{(c)} partial density of states (PDOS) of Pd($s, p, d$) orbitals in PdN$_2$;
 and \textbf{(d)} PDOS of N($s, p$) orbitals in PdN$_2$.}
\end{figure*}

The DFT(GGA) calculated electronic band structures for PdN(B17), PdN(B24) and PdN$_{2}$(C18) and their corresponding total and partial DOS are displayed in Figs. \ref{Pd1N1_B17_electronic_structure}, \ref{Pd1N1_B24_electronic_structure} and \ref{Pd1N2_C18_electronic_structure}, respectively. All phases show clear metallic feature, though PdN$_{2}$(C18) has a very low TDOS around Fermi level $E_{F}$ coming mainly from the $d$ states of the Pd atoms.

%
\subsection{\label{Optical Properties}Optical Properties}
\begin{figure*}	
\includegraphics[width=1.0\textwidth]{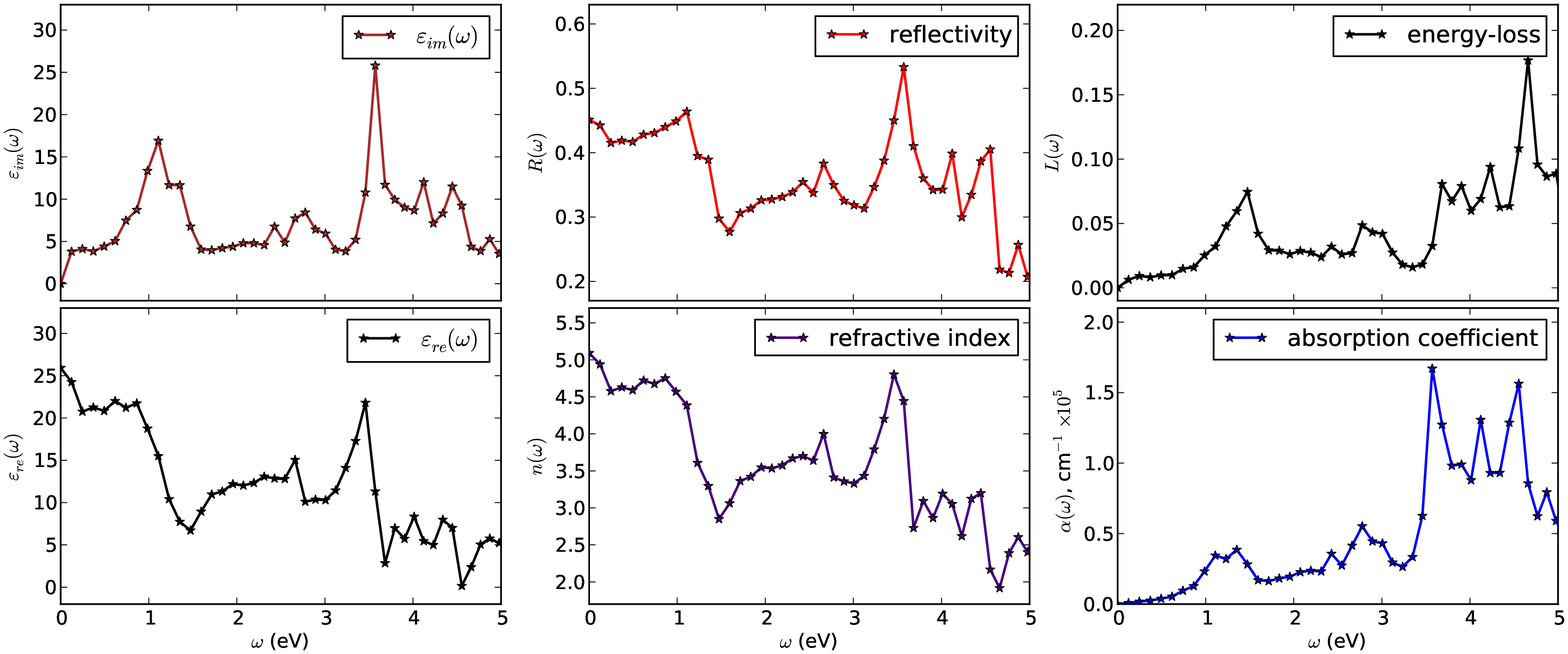}
\caption{\label{Pd1N1_B24_optical_constants}(Color online.) 
The $GW$ calculated frequency-dependent optical spectra of PdN(B24):
 \textbf{(a)} the real $\varepsilon_{\text{re}}(\omega)$ and the imaginary $\varepsilon_{\text{im}}(\omega)$ parts of the dielectric function $\varepsilon_{\text{RPA}}(\omega)$;
 \textbf{(b)} reflectivity $R(\omega)$ and transmitivity $T(\omega)$;
 \textbf{(c)} refraction $n(\omega)$ and extinction $\kappa(\omega)$ coefficients; and 
 \textbf{(d)} absorption coefficient $\alpha(\omega)$.
 The shaded area highlights the optical region.}
\end{figure*}

Fig. \ref{Pd1N1_B24_optical_constants} displays the real and the imaginary parts of $\varepsilon_{_{\text{RPA}}}(\omega)$ for PdN(B24) and the corresponding derived optical constants within the optical region $\left[\sim (3.183 - 1.655) \; eV  \equiv  (390 - 750) \; nm\right]$. With its non-zero value, it is clear from the absorption coefficient $\alpha\left(\omega\right)$ spectrum that our $G_0W_{0}$ calculations confirm that B24 is a metallic phase of PdN.

%
%
\section{\label{Concluding Remarks}Concluding Remarks}
We have applied first-principles methods to investigate the structural, electronic and optical properties of some possible stoichiometries and crystal structures of the recently discovered palladium nitride.

From the study of the equation of state (EOS), we identified the energetically most stable phases and determined their equilibrium structural parameters. B17 and C18 were found to be the most energetically favored structures in the PdN and PdN$_2$ series, respectively. Band diagrams and total and partial density of states reveal that PdN(B17 and B24) and PdN$_2$(C18) are all metallic.

The considerable differences found with and among the earlier reported structural properties (of PdN and PdN$_2$) may invoke the need for deeper and more expensive calculation schemes such as in Ref. \onlinecite{PdN2_2010_exp_n_comp}.

The more sophisticated GW approach was employed to investigate excitation energies and optical properties of this promising material. The obtained absorption coefficient spectrum confirmed that the high-pressure competing phase PdN(B24) is metallic.

In the present investigation, we have studied a wider parameter sub-space than previous works, and to the best of our knowledge, the present study is the first to propose and to investigate the physical properties of Pd$_3$N. If synthesized, Pd$_3$N will likely be in the $\epsilon$-Fe$_3$N hexagonal structure of Ni$_3$N. This Pd$_3$N modification is thermodynamically more stable (and thus it is more possible to be synthesized) than all the previously proposed PdN and PdN$_2$ modifications, and has better cohesive energy than all the previously proposed PdN and PdN$_2$(C1) modifications. Compared to all these PdN and PdN$_2$ modifications, Pd$_3$N($\epsilon$-Fe$_3$N) has the shortest Pd-Pd bond length. Moreover, Pd$_3$N($\epsilon$-Fe$_3$N) possesses slightly higher bulk modulus $B_0$ than its parent Pd, and $B_0$ increases significantly under pressure. Pd$_3$N($\epsilon$-Fe$_3$N) preserves the metallic character of its parent Pd. These properties together may make this phase, if synthesized, important to possible high-pressure applications.\\
%
\section*{Acknowledgments}
All GW calculations and some DFT calculations were carried out using the infrastructure of the Centre for High Performance Computing (CHPC) in Cape Town. Suleiman would like to acknowledge the support he received from Wits, DAAD, AIMS, SUST and the ASESMA group.
%
\bibliography{arXiv_palladium_nitrides_article}
\end{document}